\documentclass[twocolumn,amssymb,amsmath,superscriptaddress,letterpaper,pra,nofootinbib]{revtex4}

\usepackage{graphicx}
\usepackage[usenames,dvipsnames]{xcolor}

\usepackage{upgreek}

\def\appName{Appendix}

\usepackage[normalem]{ulem}

\usepackage{tikz}
\pgfdeclarelayer{edgelayer}
\pgfdeclarelayer{nodelayer}
\pgfsetlayers{edgelayer,nodelayer,main}

\tikzstyle{dot}=[circle, draw=black, fill=black!50, inner sep=1.25pt]
\tikzstyle{red}=[circle, fill=Salmon, inner sep=3pt]
\tikzstyle{green}=[circle, fill=LimeGreen, inner sep=3pt]
\tikzstyle{blue}=[circle, fill=Aquamarine, inner sep=3pt]
\tikzstyle{thick}=[line width = 0.5mm]
\tikzstyle{thin}=[line width = 0.3mm, gray]
\tikzstyle{green}=[circle, fill=LimeGreen, inner sep=5pt]

\newcommand{\ig}[1]{}

\newcommand{\ket}[1]{| #1 \rangle}
\newcommand{\bra}[1]{\langle #1 |}
\newcommand{\avg}[1]{\langle #1 \rangle}

\def\({\left(}
\def\){\right)}

\def\is{{\rm Ising}}

\newcommand{\bes} {\begin{subequations}}
\newcommand{\ees} {\end{subequations}}
\newcommand{\bea} {\begin{eqnarray}}
\newcommand{\eea} {\end{eqnarray}}
\newcommand{\beq} {\begin{equation}}
\newcommand{\eeq} {\end{equation}}

\begin{document}

\title{Experimental signature of programmable quantum annealing}

\author{Sergio Boixo}
\affiliation{Information Sciences Institute, University of Southern California}
\affiliation{Ming-Hsieh Department of Electrical Engineering}
\affiliation{Center for Quantum  Information Science \& Technology}

\author{Tameem Albash}
\affiliation{Center for Quantum  Information Science \& Technology}
\affiliation{Department of Physics and Astronomy}

\author{Federico M. Spedalieri}
\affiliation{Information Sciences Institute, University of Southern California}
\affiliation{Center for Quantum  Information Science \& Technology}

\author{Nicholas Chancellor}
\affiliation{Center for Quantum  Information Science \& Technology}
\affiliation{Department of Physics and Astronomy}

\author{Daniel A. Lidar}
\affiliation{Ming-Hsieh Department of Electrical Engineering}
\affiliation{Center for Quantum  Information Science \& Technology}
\affiliation{Department of Physics and Astronomy}
\affiliation{Department of Chemistry\\University of Southern California, Los Angeles, California
90089, USA}

\date{December 7, 2012}

\begin{abstract}

Quantum annealing is a general strategy for solving difficult optimization problems with the aid of quantum adiabatic evolution~\cite{finnila_quantum_1994,kadowaki_quantum_1998}. Both analytical and numerical evidence suggests that under idealized, closed system conditions, quantum annealing can outperform classical thermalization-based algorithms such as simulated annealing~\cite{santoro_theory_2002,morita:125210}. Do engineered quantum annealing devices effectively perform classical thermalization when coupled to a decohering thermal environment? To address this we establish, using superconducting flux qubits with programmable spin-spin couplings, an experimental signature which is consistent with quantum annealing, and at the same time inconsistent with classical thermalization, in spite of a decoherence timescale which is orders of magnitude shorter than the adiabatic evolution time.  
This suggests that programmable quantum devices, scalable with current superconducting technology, implement quantum annealing with a surprising robustness against noise and imperfections.
\end{abstract}

\maketitle 

Many optimization problems can be naturally expressed as the NP-hard problem of finding the ground state, or minimum energy configuration, of an Ising spin glass model\cite{barahona_computational_1982,Nishimori-book},
\begin{align}\label{eq:ising}
  H_\is = - \sum_{j=1}^N h_j \sigma_j^z - \sum_{1\leq j<k}^N J_{jk} \sigma_j^z \sigma_k^z\;,
\end{align}
where the parameters $h_j$ and $J_{jk}$ are,
respectively,
local fields and couplings. The operators $\sigma_j^z$ are Pauli
matrices which assign values $\{\pm 1\}$ to spin values
$\{\uparrow,\downarrow\}$.  Two algorithmic approaches designed to
address this family of problems are directly inspired by different
physical processes: classical simulated annealing (SA), and quantum
annealing (QA).

SA~\cite{kirkpatrick_optimization_1983} probabilistically explores the spin configuration space by taking into account 
the relative configuration energies and a time-dependent (fictitious)
temperature. The initial temperature is high relative to the system energy scale, to induce thermal fluctuations which prevent the system from getting trapped in
local minima. As the temperature is
lowered, the simulation is driven towards optimal solutions, represented by the global minima of the energy function. 

In QA~\cite{finnila_quantum_1994,kadowaki_quantum_1998} the dynamics are driven
by \emph{quantum}, rather than thermal fluctuations. A system implementing QA~\cite{brooke_quantum_1999,brooke_tunable_2001,RevModPhys.80.1061}
is described, at the beginning of a computation, by a transverse
  magnetic field
\begin{align}
H_{\textrm{trans}} = -\sum_{j=1}^N \sigma_j^x\;.
\end{align}
The system is initialized, at low temperature, in the ground state
of $H_{\textrm{trans}}$, an equal superposition of all $2^N$
computational basis states, the quantum analog of the initial
high-temperature classical state. The final Hamiltonian of the
computation is the function to be minimized, $H_\is$.  During the
computation, the Hamiltonian is evolved smoothly from
$H_{\textrm{trans}}$ to $H_\is$,
\begin{align}
  H(t) = A(t)H_{\textrm{trans}} + B(t) H_\is ,\quad t\in[0,T],
  \label{eq:H(t)}
\end{align}
where the ``annealing schedule" satisfies $A(0),B(T) >0$ and
$A(T)=B(0)=0$. If the change is sufficiently slow, the adiabatic
theorem of quantum mechanics predicts that the system will remain in
its ground state, and an optimal solution is
obtained~\cite{farhi_quantum_2001,Boixo:2010fk}. Similar
transformations with more general Hamiltonians are equivalent in
computational power (up to polynomial overhead) to the standard
circuit model of quantum computation~\cite{aharonov_adiabatic_2007,PhysRevLett.99.070502}, and offer at
least a quadratic speed-up over any classical SA
algorithm~\cite{somma_quantum_2008}.

Realistically, one should include the effects of coupling to a thermal
environment, i.e., consider open system quantum adiabatic
evolution~\cite{childs_robustness_2001,PhysRevLett.95.250503,amin_thermally_2008,patane_adiabatic_2008,
  de_vega_effects_2010,2012arXiv1206.4197A}.  An implementation of
open system QA has recently been reported in a programmable
architecture of superconducting flux
qubits~\cite{Harris:2010kx,Berkley:2010zr,Johnson:2010ys,johnson_quantum_2011},
and applied to relatively simple protein folding and number theory
problems~\cite{proteins-dwave,Ramsey-expt}. Although quantum tunneling
has already been demonstrated~\cite{johnson_quantum_2011}, the
decoherence time in this architecture can be three orders of magnitude
faster than the computational timescale, due in part to the
constraints imposed by the scalable design. In the circuit model of
quantum computation this relatively short decoherence time would
imply, without quantum error
correction~\cite{shor_scheme_1995,chiaverini_realization_2004}, that
the system dynamics can be described by classical
laws~\cite{Unruh:1995fk}. In the context of open system QA, this might
lead one to believe that the experimental results should be explained
by classical thermalization, or that in essence QA has effectively
degraded into SA.

\begin{figure}[t!]
\begin{center}
\begin{tikzpicture}	
  \begin{pgfonlayer}{nodelayer}
		\node [style=red] (0) at (0, 2.5) {5};
                \node [right] at (0.3, 2.5) {$-1$};
		\node [style=blue] (1) at (0, 1) {1};
                \node [right] at (0.3,1) {+1};
		\node [style=blue] (2) at (0, -1) {3};
		\node [style=blue] (3) at (-1, 0) {2};
		\node [style=blue] (4) at (1, 0) {4};
		\node [style=red] (5) at (2.5, 0) {8};
		\node [style=red] (6) at (0, -2.5) {7};
		\node [style=red] (7) at (-2.5, 0) {6};                
                \node at (-.4,1.7) {+1};
	\end{pgfonlayer}
	\begin{pgfonlayer}{edgelayer}
		\draw [style=thick] (3) to (1);
		\draw [style=thick] (3) to (2);
		\draw [style=thick] (2) to (4);
		\draw [style=thick] (4) to (1);
		\draw [style=thick] (4) to (5);
		\draw [style=thick] (0) to (1);
		\draw [style=thick] (7) to (3);
		\draw [style=thick] (6) to (2);
	\end{pgfonlayer}
\end{tikzpicture}
\end{center}
\caption{Connectivity graph of the degenerate Ising Hamiltonian used in our experiments. The four spins in the central square
  are the ``core spins" [the first four in Eq.~\eqref{eq:17states}], the four peripheral spins are
``ancillae spins" [the last four in Eq.~\eqref{eq:17states}]. As depicted, the local fields $h_j$ of the
core spins have value $+1$, the local fields of the ancillae spins have value $-1$, and all couplings $J_{jk}$ are ferromagnetic with value $1$.}
\label{fig:degH}
\end{figure}
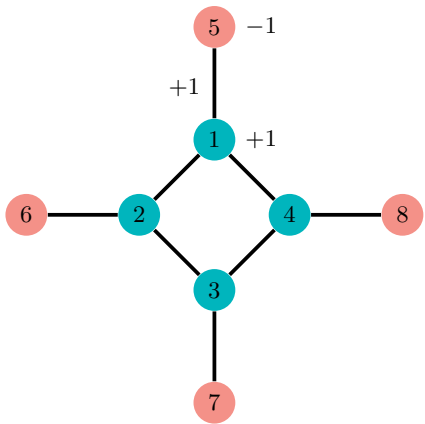

Here we address precisely this question: are the dynamics in open
system QA dominated by classical thermalization with respect to the
final Hamiltonian, as in SA, or by the energy spectrum of the
time-dependent quantum Hamiltonian?  We answer this by studying an
eight-qubit Hamiltonian representing a simple optimization problem,
and show that classical thermalization and QA make opposite
predictions about the final measurement statistics. Our Ising
Hamiltonian, depicted in Fig.~\ref{fig:degH}, has a $17$-fold
degenerate ground state \bes
\label{eq:17states}
\begin{align}
\label{eq:17states-a}
    &\{\ket{\!\!\uparrow\uparrow\uparrow\uparrow \downarrow \downarrow \downarrow \downarrow}, \dots,\ket{\!\!\uparrow\uparrow\uparrow\uparrow \downarrow \uparrow \downarrow \downarrow},\dots,\ket{\!\!\uparrow\uparrow\uparrow\uparrow \uparrow \uparrow\uparrow\uparrow}\}\\
\label{eq:17states-b}
    &\ket{\!\!\downarrow\downarrow\downarrow\downarrow\downarrow\downarrow\downarrow\downarrow}\ ,
\end{align} 
\ees Sixteen of these states form a cluster of solutions connected by
single spin-flips of the ancillae spins [Eq.~\eqref{eq:17states-a}],
while the seventeenth ground state is isolated from this cluster in
the sense that it can be reached only after at least four spin-flips
of the core spins [Eq.~\eqref{eq:17states-b}]. As we show below,
classical thermalization predicts that the isolated solution will be
found with higher probability than any of the cluster solutions, i.e.,
it is \emph{enhanced}.  Furthermore, after an initial transient,
faster thermalization corresponds to a higher probability of finding the
isolated solution. Open system QA makes the exact opposite prediction:
after an initial transient, the isolated solution is \emph{suppressed} relative to the cluster,
and faster quantum dynamics yields higher suppression (lower
probability). Our experimental results are
consistent with the open system QA prediction of the \emph{suppression
  effect}, and inconsistent with classical thermalization. We next
discuss these opposite effects, starting from the classical case.

Let $p_i$ denote the probability of state $i$ in the
cluster~\eqref{eq:17states-a}, and $p_s$ the probability of the
isolated state~\eqref{eq:17states-b}. The thermalization dynamics are
dominated by single spin-flips in our experiment (see
\appName). The probabilities $p_i$ are all close
because states in the cluster are connected by single spin flips, so
we consider the average cluster probability $p_C = \sum_{i=1}^{16}
p_i/16$. Enhancement of the isolated state means that $p_s \ge
p_C$. Our SA numerics show that this is indeed the case 
for different update rules and cooling schedules, throughout the
thermalization evolution (see \appName). To explain this, note that
the general features of a thermalization process are determined by the spectrum of $H_\is$ and by the combinatorics of state interconversion. 
Each of the $17$ degenerate ground states can be reached from any
other state without ever raising the energy via a sequence of single
spin-flips, so that SA never gets trapped in local minima
(\appName). 

The SA master equation and the classical thermalization prediction $p_s \geq p_C$ can be derived from first principles from an adiabatic quantum master equation~\cite{2012arXiv1206.4197A}. 
Let $H_S(t)$ and $H_{SB} = \sum_\alpha A_\alpha
\otimes B_\alpha$ denote the system and system-bath Hamiltonians. 
The Lindblad equation is
\begin{align}
  \dot{\rho} &= - i \left[ H_S, \rho \right] \\ \nonumber &+ \sum_{\alpha \beta}
  \sum_{a {\neq} b} \gamma_{\alpha \beta}(\omega_{ab}) \left[ L_{ab,
      \beta} \rho L^\dagger_{a b, \alpha} - \frac{1}{2} \left\{
      L^\dagger_{a b, \alpha} L_{a b,\beta}, \rho \right\}\right] \\
  \nonumber & \quad+
  \sum_{\alpha \beta} \sum_{ab} \gamma_{\alpha \beta}(0) \left[ L_{aa,
      \beta} \rho L^\dagger_{b b, \alpha} - \frac{1}{2} \left\{
      L^\dagger_{a a, \alpha} L_{b b, \beta}, \rho \right\} \right]\label{eq:2}\;,
\end{align}
where
$L_{ab,\alpha} = \ket a \bra a A_\alpha \ket b\bra b$, 
$\omega_{ab} = E_b - E_a$,
$\{\ket a\}$ is the instantaneous eigenbasis of $H_S$ for spin vector $a$, and
$\gamma_{\alpha \beta}(\omega) = \int_{-\infty}^\infty d \tau e^{i \omega\tau} \avg{B_\alpha^\dagger(\tau )B_\beta(0)}$.
We are interested in the thermalization process in which the density
operator is diagonal in the computational basis of spin vectors. The
system-bath coupling Hamiltonian then has the form 
$H_{SB} = \sum_{r\in\{+,-,z\}}\sum_{j=1}^N g^{(r)}_j \sigma_j^r\otimes B_j^{(r)}  $, where $\sigma^\pm = (\sigma^x\pm i\sigma^y)/2$.  
We denote by $a_j^+$ ($a_j^-$) the spin vector resulting from flipping the $j$th spin up (down). 
From here we arrive at the classical master equation for the populations $p_a \equiv \rho_{aa}$:
\beq
\label{eq:sa}
    \dot{p}_{a} = \sum_{j=1}^N \sum_{r=\pm}   \left(f_j(E_{a^r_j} - E_a) p_{a^r_j}- f_j(E_a-E_{a^r_j})  p_{a} \right),
\eeq
and the detailed balance condition 
$f(E_a-E_{a^\pm_j}) = \exp[-\beta (E_{a^{\pm}_ j} - E_a)]f(E_{a^\pm_j}-E_a)$.
Eq.~\eqref{eq:sa} is the master equation that we used in our SA
numerics. It can also be used to derive the classical thermalization prediction $p_s \geq p_C$. To this end, it can be seen directly that the isolated state is connected via single spin-flips to $8$ excited states with energy $-4$, giving the rate equation
$\dot p_s = 8 f(-4) p_e - 8 f(4) p_s$. In contrast, all states in the cluster are connected via single spin-flips to at most $4$ singly-excited states; the other four spin-flips connect between other states in the cluster and hence conserve the energy. Thus
$\dot{p}_C\approx 2 \( f(-4)\, p_e - f(4) \, p_C\)$.
Comparing, we
conclude that population feeds into the isolated state faster than into the cluster and,
given that initially $p_s \ge p_C$, we always
observe $p_s \ge p_C$. Simultaneous double spin-flips do not change this conclusion, and higher order simultaneous spin-flips are physically less likely. A complete derivation is given in the \appName.  

\begin{figure}[t!]
\begin{center}
\includegraphics [width= \columnwidth]{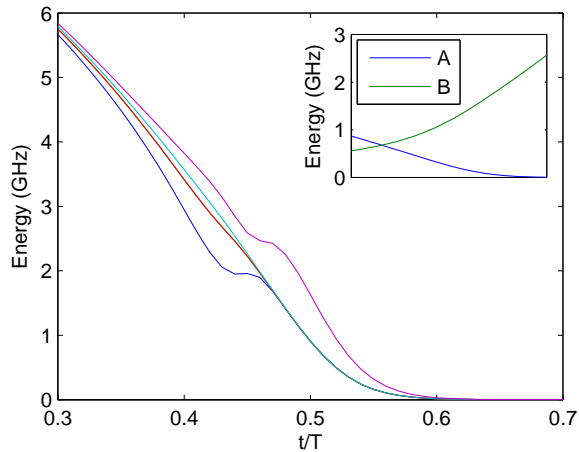}
\end{center}
\caption{The time-dependent gap between the ground state and the lowest six excited states in the relevant region of the experimental QA   evolution. After time $t = 0.5T$ the highest energy level shown corresponds to the isolated state. The inset shows the transverse field magnitude $A(t)$ and Ising Hamiltonian magnitude $B(t)$ used in our experiments, during the same time interval.}
\label{fig:gaps}
\end{figure}


We next analyze the corresponding predictions of QA.
A crucial difference with respect to SA is that now the relevant energy spectrum is given by a combination of
the final Ising Hamiltonian and the transverse field. Consequently, as shown in Fig.~\ref{fig:gaps},
the degeneracy of the ground space is lifted for times $t<T$. The
isolated state has support only on the highest eigenstate plotted
during the second half of the evolution.  Given that the system starts
in the ground state, the isolated state is suppressed by the energy
gap, until this gap vanishes at the end of the evolution. The isolated
state remains suppressed nonetheless, since transitions to other low
energy states require at least four spins-flips. The transverse field
term, which drives simultaneous spin-flip transitions, is small at
large $t$. If the four spins-flips are not simultaneous, these
transitions involve excited states with much higher energy, and are
suppressed. This predicted QA suppression of the isolated state is
confirmed by our closed and open system quantum dynamical simulations.

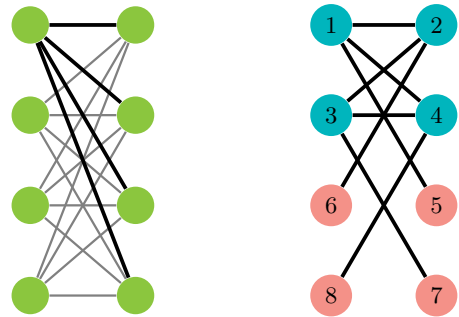
\begin{figure}
\centering
\begin{minipage}{0.45\columnwidth}
\centering
\begin{tikzpicture}	
  \begin{pgfonlayer}{nodelayer}
		\node [style=green] (0) at (0, 3.6) {};
		\node [style=green] (1) at (0, 2.4) {};
		\node [style=green] (2) at (0, 1.2) {};
		\node [style=green] (3) at (0, 0) {};
		\node [style=green] (4) at (1.4, 3.6) {};
		\node [style=green] (5) at (1.4, 2.4) {};
		\node [style=green] (6) at (1.4, 1.2) {};
		\node [style=green] (7) at (1.4, 0) {};  
		\draw [style=thick] (0) to (4);
		\draw [style=thick] (0) to (5);
		\draw [style=thick] (0) to (6);
		\draw [style=thick] (0) to (7);
	\end{pgfonlayer}
	\begin{pgfonlayer}{edgelayer}
		\draw [style=thin] (1) to (4);
		\draw [style=thin] (1) to (5);
		\draw [style=thin] (1) to (6);
		\draw [style=thin] (1) to (7);
		\draw [style=thin] (2) to (4);
		\draw [style=thin] (2) to (5);
		\draw [style=thin] (2) to (6);
		\draw [style=thin] (2) to (7);
		\draw [style=thin] (3) to (4);
		\draw [style=thin] (3) to (5);
		\draw [style=thin] (3) to (6);
		\draw [style=thin] (3) to (7);
	\end{pgfonlayer}
\end{tikzpicture}
\end{minipage}
\begin{minipage}{0.45\columnwidth}
\centering
\begin{tikzpicture}	
  \begin{pgfonlayer}{nodelayer}
		\node [style=blue] (0) at (0, 3.6) {1};
		\node [style=blue] (1) at (0, 2.4) {3};
		\node [style=red] (2) at (0, 1.2) {6};
		\node [style=red] (3) at (0, 0) {8};
		\node [style=blue] (4) at (1.4, 3.6) {2};
		\node [style=blue] (5) at (1.4, 2.4) {4};
		\node [style=red] (6) at (1.4, 1.2) {5};
		\node [style=red] (7) at (1.4, 0) {7};  
	\end{pgfonlayer}
	\begin{pgfonlayer}{edgelayer}
		\draw [style=thick] (0) to (4);
		\draw [style=thick] (0) to (5);
		\draw [style=thick] (0) to (6);
		\draw [style=thick] (1) to (4);
		\draw [style=thick] (1) to (5);
		\draw [style=thick] (1) to (7);
		\draw [style=thick] (2) to (4);
		\draw [style=thick] (3) to (5);
	\end{pgfonlayer}
\end{tikzpicture}
\end{minipage}
\caption{Left: schematic depicting the maximal connectivity graph ($K_{4,4}$) of the qubits inside a unit cell. Right: an embedding of $H_\is$ from Fig.~\ref{fig:degH} (right).}
\label{fig:layouts}
\end{figure}

Our experiments were performed using the D-Wave One Rainier chip at the USC Information Sciences Institute, comprising $16$ unit cells of $8$ superconducting flux qubits each, with a total of $108$ functional qubits. The couplings are programmable
superconducting inductances. The qubits and unit cell, readout, and control have been described in detail elsewhere~\cite{Harris:2010kx,Berkley:2010zr,Johnson:2010ys,johnson_quantum_2011}. The initial energy scale for the transverse field is $10$GHz (the $A$ function in Fig~\ref{fig:gaps}), the final
energy scale for the Ising Hamiltonian (the $B$ function) is $5.3$GHz, about fifteen times the experimental temperature of $17$mK $\approx 0.35$GHz. To gather our data, we ran each of the $144$ embeddings $4000$ times, in batches of
$1000$ readouts, resetting all the local fields and couplers after each batch.

\begin{figure*}[t!]
\begin{center}
\includegraphics [width= \textwidth]{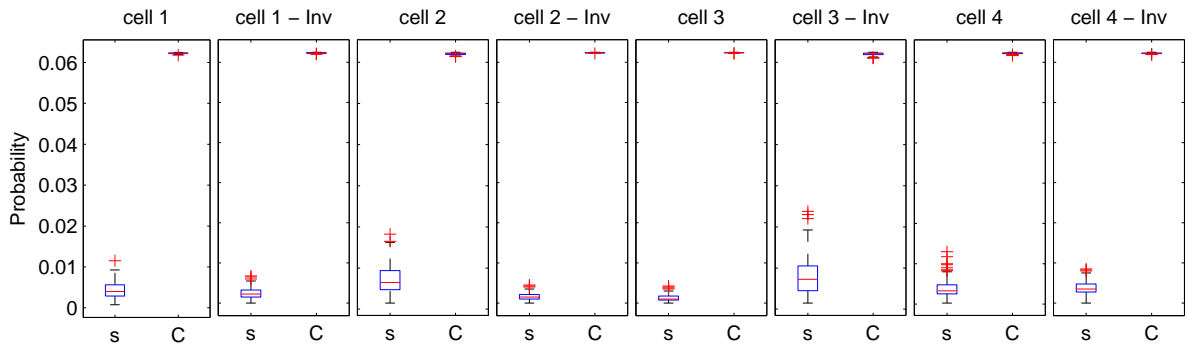}
\end{center}
\caption{Statistical box plot~\cite{Frigge:1989vn} 
of the experimental results for $p_s$ (left columns) and $p_C$ (right columns). The total annealing time was $T=5 \mu$s in each run. In each column the bar is the median, the box corresponds to the lower and upper quartiles, 
respectively, 
the segment contains most of the samples, and the $+$'s are outliers. 
Cells 1-4 are physically distinct $8$-qubit unit cells on the chip. No statistically significant variation is seen as a function of the unit cell number. ``Inv" is the case where each local field $h_j$ is flipped to $-h_j$. In this case the isolated state corresponds to the state $\ket{\!\!\uparrow\uparrow\uparrow\uparrow\uparrow\uparrow\uparrow\uparrow}$, the opposite of Eq.~\eqref{eq:17states-b}. While this has a small effect for cells $2$ and $3$, in all cases the isolated state is significantly suppressed, as predicted by QA. This establishes that suppression of the isolated state is not due to a global magnetic field bias.
}
\label{fig:results5us}
\end{figure*}

A diagram of the experimentally achievable coupling configurations is shown in
Fig.~\ref{fig:layouts} (left). 
The experimental results are shown in
Fig.~\ref{fig:results5us}. The key finding that is immediately apparent is that the isolated state is robustly suppressed, in agreement with the QA but not the SA prediction. 

Is it possible that suppression has an explanation other than QA? The
main physical argument along these lines is that a systematic or
random bias due to experimental imperfections breaks the $17$-fold
ground state degeneracy and energetically disfavors the isolated
state, thus lowering $p_s$ if the system thermalizes. We proceed to
examine this and the robustness of the suppression effect.

First, note that spin numbers $j=1,\dots,8$ must be assigned to the
flux qubits before each experimental run. One of the $144$ possible
such ``embeddings" allowed by the symmetries of the Hamiltonian and
the hardware connectivity-graph is shown in
Fig.~\ref{fig:layouts} (right).

\begin{figure*}[t!]
\begin{center}
\includegraphics [width= \textwidth]{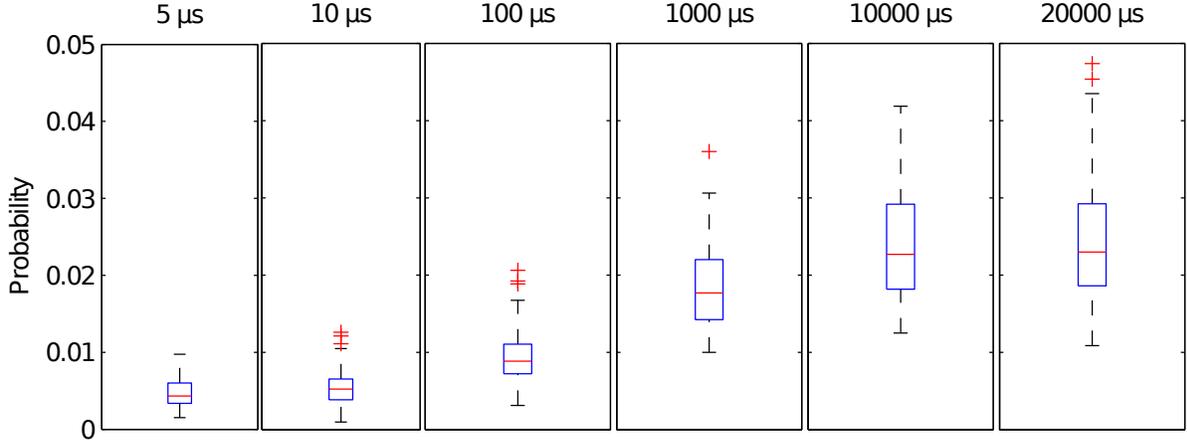}
\end{center}
\caption{Statistical box plot of the probability of the isolated state for a fixed set of qubits, with different annealing times. In this plot each of the $144$ possible embeddings is averaged with the same embedding after a complete spin inversion. This compensates for global magnetic field biases (which can be seen in Fig.~\ref{fig:results5us}, cells 2 and 3). The Ising Hamiltonian for this data was obtained by applying a random spin inversion to $H_\is$ from Fig.~\ref{fig:degH}. The probability of the isolated state increases with the annealing time $T$, in contrast to the classical
thermalization prediction. While in classical thermalization an initial distribution concentrated around the cluster can also result in suppression
of the isolated state, this suppression is highly
unlikely to persist after a random inversion.}
\label{fig:overTime}
\end{figure*}

Second, note that spin-inversion transformations $H(t) \to \sigma_j^x
H(t) \sigma_j^x$ commute with $H_{\textrm{trans}}$, and simply relabel
the spectrum of both $H_\is$ and $H(t)$: if a certain spin
configuration has energy $E$, then the corresponding spin
configuration with the $j$th spin flipped has the same energy $E$
under $\sigma_j^x H_\is \sigma_j^x$.
Spin-inversions also commute with the spin-flip operations of
classical thermalization. Therefore all of our arguments for the
suppression of the isolated state in QA and for its enhancement in
classical thermalization are unchanged.  Using spin inversions we can
check that the suppression effect is not due to a perturbation of the
Hamiltonian such as a magnetic field bias.
Indeed, by performing a spin-inversion on all $N$ spins we obtain a new Ising Hamiltonian where the isolated state is that with all spins-up. If a field bias suppressed the all spins-down state, then it would enhance the all spins-up state. Figure~\ref{fig:results5us} rules this out.
We also tested cases with only antiferromagnetic couplings, and with
random spin-inversions. The results for one such random inversion
example are shown in Fig.~\ref{fig:overTime}. In all cases we found
agreement with the QA prediction, but not with classical
thermalization. 

Robust suppression holds even at the level of individual embeddings
and spin inversions. We found that $p_s\lesssim 3$\%, while
$p_C\gtrsim 6$\% for each of the thousands of such cases we tested
(the highest median for the experiment in Fig~\ref{fig:results5us} is
0.004). Thus suppression survives breaking of the ground state
degeneracy, which certainly occurs due to the limited precision of
$\sim 5$\% in our control of $\{h_j,J_{jk}\}$. The suppression effect
is robust because it does not depend on the exact values of these
parameters, but on the relatively large Hamming distance between the
isolated state and the cluster.

\begin{figure}[t!]
\begin{center}
\includegraphics [width= \columnwidth]{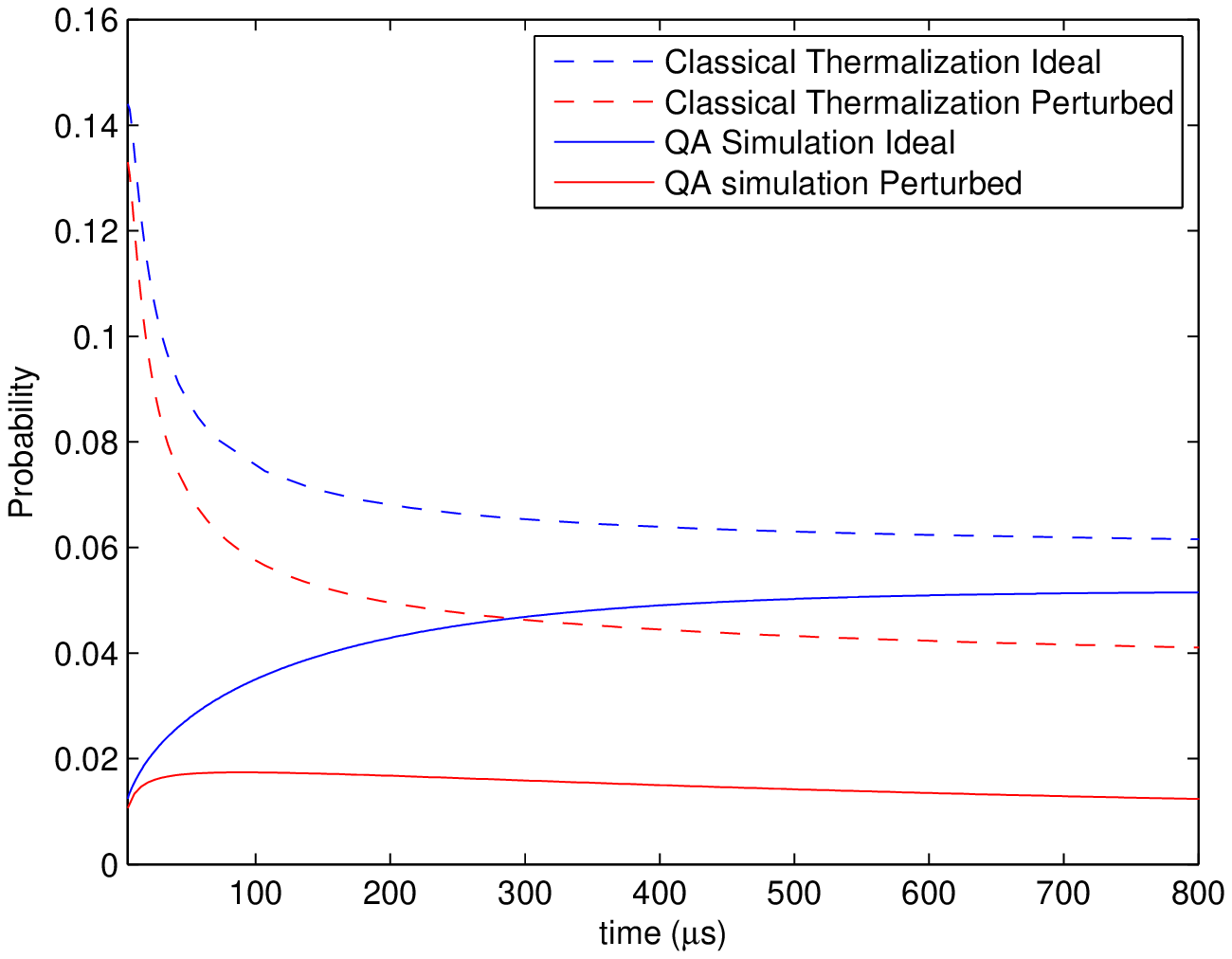}
\end{center}
\caption{Probability of the isolated state for numerical simulations of classical thermalization (Metropolis update rule) and open system QA as a function of the total annealing time $T$. ``Ideal" \textit{vs} ``perturbed" corresponds to simulations for $H_\is$ without and with a perturbation which increases the
energy of the isolated state (red). In classical thermalization $p_s$ always decreases with $T$, while it increases for QA in the ideal case. It remains almost constant for QA with the perturbed Hamiltonian. 
Even if the isolated state is suppressed energetically due to a perturbation of $H_\is$, fast classical
thermalization can still enhance its probability.  QA with the ideal Hamiltonian gives the best qualitative fit to the experimental data. System-bath coupling in the QA simulation corresponds to a decoherence time of $150$ns.}
\label{fig:SAvsQA}
\end{figure}

Finally, we consider the effect of increasing the annealing time.
Open quantum and classical systems converge towards thermal
equilibrium. Therefore if the cause of suppression is the QA spectrum,
longer annealing times will result in $p_s$ \emph{increasing},
approaching its Gibbs distribution value. This would not be the case
if $p_s$ were governed by the spectrum of $H_\is$. In
Fig.~\ref{fig:SAvsQA} we compare a numerical simulation of open system
QA, using an adiabatic Markovian master
equation~\cite{2012arXiv1206.4197A}, with classical thermalization. The
quantum prediction of increasing $p_s$ is confirmed experimentally, as
shown in Fig.~\ref{fig:overTime}.

We thus arrive at our main conclusion: signatures of QA, as opposed to
classical thermalization, persist for timescales that are much longer
than the single-qubit decoherence time (from $5\mu$s to $20$ms
\textit{vs} tens of ns) in programmable devices available with
present-day superconducting technology. Our experimental results are
also consistent with numerical methods that compute quantum
statistics, such as Path Integral Monte Carlo~\footnote{M. Troyer and
T. Roennow, private communication.}. Our study focuses on demonstrating
a non-classical signature in experimental QA.  Different methods are
required to address the question of experimental computational
speedups of open system QA relative to optimal classical algorithms.
\vspace{.5cm}

\paragraph*{Acknowledgements.}
We thank M. Amin, T. Landing, T. Murray, J. Preskill, T. Roennow and
M. Troyer for useful discussions. We particularly thank M. Amin for
discussions that inspired our choice of the Ising Hamiltonian. This
research was supported by the Lockheed Martin Corporation. S.B. and
D.A.L. acknowledge support under ARO grant number
W911NF-12-1-0523. D.A.L. was further supported by the National Science
Foundation under grant number CHM-1037992, and ARO MURI grant
W911NF-11-1-0268.

\appendix

\section{First-principles derivation of the master equation}

Here we derive the master equation used by SA from
first principles within the open quantum systems formalism. This motivates classical SA as a model for a system dominated by classical thermalization of the final Ising Hamiltonian.

Let $H_S(t)$ be the time-dependent system Hamiltonian and $H_{SB} = \sum_\alpha A_\alpha
\otimes B_\alpha$ be the system-bath Hamiltonian. 
We have previously established that the Lindblad equation within the rotating wave approximation has the form~\cite{2012arXiv1206.4197A}
\begin{align}  
\label{eq:H_S+H_LS}
&\dot{\rho} = - i \left[ H_S, \rho \right] \\
&\; +
\sum_{\alpha \beta} \sum_{a {\neq} b} \gamma_{\alpha^\star \beta}(\omega_{ab}) \left[
L_{ab, \beta} \rho L^\dagger_{a b, \alpha}  - \frac{1}{2} \left\{
  L^\dagger_{a b, \alpha} L_{a b,\beta}, \rho \right\}\right]
\nonumber \\
&\;+ \sum_{\alpha \beta} \sum_{ab} \gamma_{\alpha^\star \beta}(0) \left[
L_{aa, \beta} \rho L^\dagger_{b b, \alpha} - \frac{1}{2} \left\{ L^\dagger_{a a, \alpha} L_{b b,
\beta}, \rho \right\} \right]  \nonumber \ ,
\end{align}
where
\bes
\begin{align}
  L_{ab,\alpha} &= \ket a \bra a A_\alpha \ket b\bra b \\
  L_{ab,\alpha}^\dagger &= \ket b \bra b A^\dagger_\alpha \ket a\bra a \\
  \label{eq:omega_ab}
  \omega_{ab} &= E_b - E_a\;,
\end{align}
\ees
$\{\ket a\}$ is the instantaneous eigenbasis of $H_S$ (we have suppressed its explicit time-dependence) for spin vector $a=\{a_1,\dots,a_N\}$, where $a_i \in \{\uparrow,\downarrow\}$ , and
\begin{align}
  \label{eq:gamma_def}
\gamma_{\alpha^\star \beta}(\omega) = \int_{-\infty}^\infty d \tau e^{i \omega
\tau} \avg{B_\alpha^\dagger(\tau )B_\beta(0)}
\end{align}
is the Fourier transform of the bath correlation function. The star adornment on the first subscript ($\alpha^\star$) is a reminder
that the first operator in the bath correlation function is Hermitian-transposed. We have ignored the Lamb shift in Eq.~\eqref{eq:H_S+H_LS}
since for a time-dependent
Lindblad evolution it amounts to a small perturbation of the system Hamiltonian.  We used this form of the master equation for our quantum open system
numerical simulations, as detailed elsewhere~\cite{2012arXiv1206.4197A}.

We show in Sec.~\ref{subsec:KMS} that for a bath in thermal equilibrium at inverse temperature $\upbeta$
\begin{align}
\gamma_{\alpha^\star \beta}(-\omega) = e^{-\upbeta\omega}\gamma_{\beta \alpha^\star}(\omega)\ ,
\label{eq:gammaflip}
\end{align}
where
\begin{align}
\label{eq:gamma_def2}
    \gamma_{\beta \alpha^\star}(\omega) = \int_{-\infty}^\infty d \tau e^{i \omega
\tau} \avg{B_\beta(\tau )B_\alpha^\dagger(0)} \;.
\end{align}

We assume that the
system-bath coupling Hamiltonian has the form
\begin{align}\label{eq:SB}
H_{SB} = \sum_{j=1}^N \sum_{r\in\{\pm,z\}} g^{(r)}_j \sigma_j^r\otimes B_j^{(r)} ,
  \end{align}
where $\sigma^\pm = (\sigma^x\pm i\sigma^y)/2$, we identify $\ket{\!\!\uparrow}$ with $\ket{0}$ and $\ket{\!\!\downarrow}$ with $\ket{1}$, and where we neglect higher-order interactions of the form $\sigma_j^r\otimes \sigma_k^s\otimes B_{jk}^{(rs)}$ or above. Since $H_{SB}$ is Hermitian we also have $B_j^{(\pm)\dagger} = B_j^{(\mp)}$, $B_j^{(z)\dagger} = B_j^{(z)}$, $g_j^{(\pm)*} = g_j^{(\mp)}$, $g_j^{(z)*} = g_j^{(z)}$, and where the asterisk denotes complex conjugation. In the computational basis of spin vectors $\{a\}$, we introduce the notation 
\beq
\ket{a^{\pm}_{ j}} \equiv \sigma_j^{\pm} \ket{a}\; ,
\eeq 
which denotes either a flipping of $a_j$, or $0$ if either $\sigma_j^+$ acts on $a_j = \uparrow$ or  $\sigma_j^-$ acts on $a_j = \downarrow$. Then 
\begin{align}
    \bra{a}\sigma^{\pm}_j\ket{b} = \(\sigma^{\pm}_j\)_{ab} = \delta_{a,b^{\pm}_{j }} \; ,
\label{eq:sigab}
\end{align}
where the $\delta$ function is defined to evaluate to zero also when $\sigma^{\pm}_j$ annihilates $\ket{b}$. We are interested in classical thermalization, in which the density
operator is diagonal in the computational basis $\{a\}$, so we set
$\rho_{ab}  = 0$ for $a \ne b$. Equation~\eqref{eq:H_S+H_LS} then
gives $\dot \rho_{ab} = 0$. Using indexes $\alpha = (r,j)$ and $\beta = (s,k)$ in Eq.~\eqref{eq:H_S+H_LS}, where $r,s\in\{\pm,z\}$ and $j,k\in [1,\dots,N]$, and taking the diagonal $\bra{a}\cdot\ket{a}$ matrix element, the Lindblad equation becomes
\begin{align}
\label{eq:rhodotaa}
    \dot{\rho}_{aa} &= \sum_{(r,j), (s,k)} g_j^{(r)*} g_k^{(s)}\times \\
&   \sum_{b|b {\neq} a}\gamma_{(r,j)^\star (s,k)}(\omega_{ab})
\(\sigma^{s}_k\)_{ab} \rho_{bb} \((\sigma^{r}_j)^\dagger\)_{ba}
\nonumber\\ &- \gamma_{(r,j)^\star (s,k)}(\omega_{ba}) 
  \((\sigma^{r}_j)^\dagger\)_{ab} \rho_{aa} \(\sigma^{s}_k\)_{ba}\nonumber \; .
\end{align}
Note that the sum in Eq.~\eqref{eq:H_S+H_LS} involving the resonant contribution $\gamma_{\alpha^\star \beta}(0)$ vanishes, since the terms $L_{aa, \beta} \rho L^\dagger_{b b, \alpha}$ and $\frac{1}{2} \left\{ L^\dagger_{a a, \alpha} L_{b b,
\beta}, \rho \right\}$ cancel after taking the diagonal matrix element. Moreover, since Eq.~\eqref{eq:rhodotaa} involves only off-diagonal terms ($b\neq a$), contributions due to $\sigma^z$ all vanish, and using Eq.~\eqref{eq:sigab}, we are left with 
\begin{align}
    \dot{p}_{a} &= \sum_{j=1}^N \sum_{r=\pm}  |g_j^{(r)}|^2 \left(\gamma_{(r,j)^\star (r,j)}(\omega_{aa^{-r}_j}) p_{a^{-r}_j}\right. \nonumber
\\ &- \left. \gamma_{(r,j)^\star (r,j)}(\omega_{a^r_ja})  p_{a} \right) \;,
\label{eq:rhodotaa1}
\end{align}
where we denoted $p_a \equiv \rho_{aa}$, the probability of spin configuration $a$. We can furthermore identify 
\bes
\label{eq:P-P}
\begin{align}
&  P(a \to a^{r}_{j}) \equiv |g_j^{(r)}|^2 \gamma_{(r,j)^\star (r,j) } (\omega_{a^{r}_{j}a})\\
& P(a^{-r}_{j} \to a) \equiv |g_j^{(r)}|^2 \gamma_{(r,j)^\star (r,j) } (\omega_{aa^{-r}_{j}})
\;
\end{align}
\ees
as the transition probabilities, so that Eq.~\eqref{eq:rhodotaa1} becomes the rate equation
\beq
\dot{p}_{a} = \sum_{j=1}^N \sum_{r=\pm}  P(a^{-r}_{j} \to a) p_{a^{-r}_j}- P(a \to a^{r}_{j})  p_{a} \;.
\label{eq:rate-eqn}
\eeq

This can be further simplified using the KMS condition. Indeed, note that, using $B_\alpha(\tau) = \sigma_j^{\pm}(\tau)$ in Eqs.~\eqref{eq:gamma_def} and \eqref{eq:gamma_def2}, we have
\beq
\gamma_{(\pm,j)^\star(\pm,j)}(\omega) = \gamma_{(\mp,j)(\mp,j)^\star}(\omega) .
\eeq
Using this along with $\omega_{a^{\pm}_{j}a} = - \omega_{aa^{\pm}_{j}}$ [Eq.~\eqref{eq:omega_ab}] and Eq.~\eqref{eq:gammaflip}, we have
\begin{align}
\gamma_{(\pm,j)^\star (\pm,j)}(\omega_{a^{\pm}_{j}a}) 
= e^{-\upbeta\omega_{aa^{\pm}_{j}}}\gamma_{(\mp,j)^\star (\mp,j)}(\omega_{aa^{\pm}_{j}})\ .
\end{align}
Therefore Eq.~\eqref{eq:P-P} yields
\bes
\label{eq:Ps}
\begin{align}
P(a \to a^{\pm}_{j}) & =  e^{-\upbeta\omega_{aa^{\pm}_{j}}}|g_j^{(\pm)}|^2\gamma_{(\mp,j)^\star (\mp,j)}(\omega_{aa^{\pm}_{j}}) \\
P(a^{\pm}_{j} \to a) & = |g_j^{(\mp)}|^2\gamma_{(\mp,j)^\star (\mp,j)}(\omega_{aa^{\pm}_{j}})\; .
\end{align}
\ees
This, together with $g_j^{(\pm)*} = g_j^{(\mp)}$, gives the detailed balance condition for
thermalization dynamics
\begin{align}
\label{eq:dbc}
\frac{P(a \to a^{\pm}_j)} { P(a^{\pm}_j \to a)} 
= e^{-\upbeta (E_{a^{\pm}_ j} - E_a)} = \frac{f_j(E_a-E_{a^\pm_j})}{f_j(E_{a^\pm_j}-E_a)}\; ,
\end{align}
where we introduced transition functions $f_j(\Delta E)$, which we identify with the transition probabilities in Eq.~\eqref{eq:Ps}.

We can now rewrite Eq.~\eqref{eq:rate-eqn} as the classical master equation that we used in our SA
numerical simulations
\begin{align}
\label{eq:sa2}
\dot{p}_{a} 
& =  \sum_{j=1}^N \sum_{r=\pm}   \left(f_j(E_{a^r_j} - E_a)p_{a^r_j}- f_j(E_a-E_{a^r_j})  p_{a} \right) \;.
\end{align}

\section{Correlation functions and the KMS condition}
\label{subsec:KMS}

Here we derive the detailed balance condition Eq.~\eqref{eq:gammaflip} from first principles. Our calculation closely follows Ref.~\onlinecite{2012arXiv1206.4197A}, but differs in that it applies also to non-Hermitian bath operators.

The correlation function of a thermal bath is assumed to satisfy the
KMS (Kubo-Martin-Schwinger) condition~\cite{Breuer:2002}
\beq
\label{eq:KMSt}
\langle B^\dagger_\alpha(\tau)B_\beta(0)\rangle = \langle
B_\beta(0)B^\dagger_\alpha(\tau+i\upbeta)\rangle \ .  
\eeq 
This expression has
the advantage that it also applies to operators which are not trace class. For trace class operators the KMS condition can be derived assuming that the bath is in a thermal state, $\rho_B = e^{-\upbeta H_B}$, where $H_B$ is the bath Hamiltonian. In this case: 
\begin{align}
  &\langle B_\alpha^\dagger(\tau)B_\beta(0)\rangle = \textrm{Tr}[\rho_B
  U^\dag_B(\tau,0) B_\alpha^\dagger U_B(\tau,0) B_\beta] \notag \\ 
  &\;= \frac{1}{\mathcal{Z}}\textrm{Tr}[B_\beta e^{-(\upbeta-i\tau)H_B} B_\alpha^\dagger e^{-i\tau H_B}] \notag \\
  &\;= \frac{1}{\mathcal{Z}}\textrm{Tr}[B_\beta e^{i(\tau+i\upbeta)H_B} B_\alpha^\dagger
  e^{-i(\tau+i\upbeta) H_B}e^{-\beta H_B}] \notag \\
  &\;= \textrm{Tr}[\rho_B B_\beta U^\dag_B(\tau+i\upbeta,0) B_\alpha^\dagger U_B(\tau+i\upbeta,0) ] \notag \\
  &\;= \langle B_\beta(0)B_\alpha^\dagger(\tau+i\upbeta)\rangle\ ,\label{eq:4}
  \end{align}
  where $U_B$ is the bath unitary evolution operator.
Note that
\begin{align}
  \langle B_\alpha^\dagger(\tau)B_\beta(0)\rangle &= \langle B_\beta(-\tau - i \upbeta)
  B^\dagger_\alpha(0) \rangle\ .
  \label{eq:5}
  \end{align}

If in addition the correlation function
is analytic in the strip between $\tau=-i\upbeta$ and $\tau=0$, then it
follows that the Fourier transform of the bath correlation function
satisfies the detailed balance condition Eq.~\eqref{eq:gammaflip} as we show next. 

\begin{figure}[h]
  \centering
   \includegraphics[width=2in]{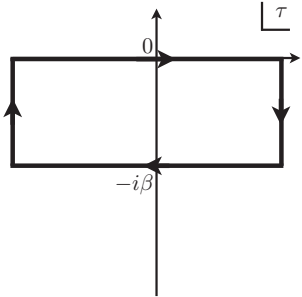} 
   \caption{Contour used in our proof of the KMS condition.}    
   \label{fig:Contour}
\end{figure}

We compute the Fourier transform:
\begin{align}
  \gamma_{\alpha^\star \beta} (\omega) &= \int_{-\infty}^{\infty} d \tau e^{i \omega
    \tau} \langle B_\alpha^\dagger(\tau)B_\beta(0)\rangle \notag \\ 
    &= \int_{-\infty}^{\infty} d
  \tau e^{i \omega \tau} \langle B_\beta(-\tau - i \upbeta) B^\dagger_\alpha(0)
  \rangle \; .
  \label{eqt:integrand1}
  \end{align}
To perform this integral we replace it with a contour integral in the complex plane, 
$\oint_C d \tau e^{i \omega \tau} \langle B_\beta(-\tau - i \upbeta) B^\dagger_\alpha(0) \rangle $,
with the contour $C$ as shown in Fig.~\ref{fig:Contour}. This contour integral vanishes by the Cauchy-Goursat theorem~\cite{complex:book} since the closed contour encloses no poles (the correlation function $\langle B_\beta(\tau) B^\dagger_\alpha(0)\rangle$ is analytic in the open strip $(0, -i \beta)$ and is continuous at the boundary of the strip~\cite{KMS}), so that
\begin{align}
\oint_{C} \left( \dots \right) &= 0 \\&= \int_{\uparrow} \left( \dots \right) + \int_{\mathrm{\downarrow}}  \left( \dots \right)  + \int_{\rightarrow}  \left( \dots \right)  +  \int_{\leftarrow}  \left( \dots \right)\; , \nonumber 
\end{align}
where $\left( \dots \right)$ is the integrand of Eq.~\eqref{eqt:integrand1}, and the integral $ \int_{\rightarrow}$ is the same as in Eq.~\eqref{eqt:integrand1}.  After making the variable transformation $\tau = -x - i \upbeta$, where $x$ is real, we have
\begin{equation}
\int_{\leftarrow}  \left( \dots \right)  = -  e^{\upbeta \omega}  \int_{-\infty}^{\infty} e^{-i \omega x}  \langle B_\beta(x) B^\dagger_\alpha(0) \rangle\; .
\end{equation}
Assuming that $\langle B_\alpha(\pm\infty-i\upbeta)B_\beta(0)\rangle = 0$ (i.e., the correlation function vanishes at infinite time), we further have $\int_{\uparrow} \left( \dots \right) = \int_{\mathrm{\downarrow}}  \left( \dots \right) =0$, and hence we find the result:
\begin{align}
  \label{eq:1}
  &\int_{-\infty}^{\infty} d \tau e^{i \omega \tau} \langle
  B_\beta(-\tau - i \upbeta) B^\dagger_\alpha(0) \rangle \\ 
  & = e^{\upbeta
    \omega} \int_{-\infty}^{\infty} e^{-i \omega \tau} \langle
  B_\beta(\tau) B^\dagger_\alpha (0) \rangle = e^{\upbeta \omega}
  \gamma_{\beta \alpha^\star}(-\omega)\ ,\notag
  \end{align}
which, together with Eq.~\eqref{eqt:integrand1}, proves
Eq.~\eqref{eq:gammaflip}. 

\section{Spectrum and ground states of the Ising Hamiltonian}

The spectrum of the $8$-qubit Ising Hamiltonian we consider in the main text, 
\begin{center}
\begin{tikzpicture}	
  \begin{pgfonlayer}{nodelayer}
		\node [style=red] (0) at (0, 2.5) {5};
                \node [right] at (0.3, 2.5) {$-1$};
		\node [style=blue] (1) at (0, 1) {1};
                \node [right] at (0.3,1) {+1};
		\node [style=blue] (2) at (0, -1) {3};
		\node [style=blue] (3) at (-1, 0) {2};
		\node [style=blue] (4) at (1, 0) {4};
		\node [style=red] (5) at (2.5, 0) {8};
		\node [style=red] (6) at (0, -2.5) {7};
		\node [style=red] (7) at (-2.5, 0) {6};                
                \node at (-.4,1.7) {+1};
	\end{pgfonlayer}
	\begin{pgfonlayer}{edgelayer}
		\draw [style=thick] (3) to (1);
		\draw [style=thick] (3) to (2);
		\draw [style=thick] (2) to (4);
		\draw [style=thick] (4) to (1);
		\draw [style=thick] (4) to (5);
		\draw [style=thick] (0) to (1);
		\draw [style=thick] (7) to (3);
		\draw [style=thick] (6) to (2);
	\end{pgfonlayer}
\end{tikzpicture}
\end{center}
can be analyzed by
first considering the spectrum of the Hamiltonian coupling a single ancilla spin to a core spin, i.e.,
\begin{center}
\begin{tikzpicture}	
  \begin{pgfonlayer}{nodelayer}
                \node  at (-1,0.6) {$-1$};
                \node  at (-2.5,0.6) {$+1$};
                \node  at (-1.7,-0.4) {$+1$};
		\node [style=red] (3) at (-1, 0) {A};
		\node [style=blue] (7) at (-2.5, 0) {C};    
	\end{pgfonlayer}
	\begin{pgfonlayer}{edgelayer}
		\draw [style=thick] (3) to (7);
	\end{pgfonlayer}
\end{tikzpicture}
\end{center}
The spectrum, with the core (ancilla) spin written first (second) is
\begin{align}
\label{2qspectrum}
  \begin{array}{r|r}
    \ket{\!\!\uparrow\uparrow} & -1 \\ \ket{\!\!\uparrow\downarrow} & -1 \\ \ket{\!\!\downarrow\uparrow} & 3 \\
    \ket{\!\!\downarrow\downarrow} & -1
  \end{array}
\end{align}
The minimum energy is $-1$ whether the core spin is up or down. It is important to note that if the core spin is up, the minimum energy is $-1$ whether the ancilla is up or down; this will give rise to a $16$-fold degeneracy when we account for all spins below.  

We analyze the core spins' energies by first taking into account only their
couplings. That is, we analyze the ferromagnetic Hamiltonian 
\begin{center}
\begin{tikzpicture}	
  \begin{pgfonlayer}{nodelayer}
		\node [style=blue] (1) at (0, 1) {1};
		\node [style=blue] (2) at (0, -1) {3};
		\node [style=blue] (3) at (-1, 0) {2};
		\node [style=blue] (4) at (1, 0) {4};
	\end{pgfonlayer}
	\begin{pgfonlayer}{edgelayer}
		\draw [style=thick] (3) to (1);
		\draw [style=thick] (3) to (2);
		\draw [style=thick] (2) to (4);
		\draw [style=thick] (4) to (1);
	\end{pgfonlayer}
\end{tikzpicture}
\end{center}
Denoting by $s$ the number of 
satisfied couplings (both spins linked by the coupling have the same sign), the
energy is $4-2s$, where $s\in\{0,2,4\}$. The ground states of this Hamiltonian are the
configurations $\ket{\!\!\uparrow\uparrow\uparrow\uparrow}$ and $\ket{\!\!\downarrow\downarrow\downarrow\downarrow}$. Since Eq.~\eqref{2qspectrum} shows that the minimum energy of a core-ancilla pair is $-1$, when adding the low
energy configurations of the couplings to the ancillae the minimum
energy is $-8$. It also follows from Eq.~\eqref{2qspectrum} that the ground state configurations of the full $8$-qubit Hamiltonian are 
\bes
\begin{align}
\label{eq:alldown}
 &\ket{\!\!\downarrow\downarrow\downarrow\downarrow\downarrow\downarrow\downarrow\downarrow\,} \\
 \label{eq:updown}   
&\ket{\!\!\uparrow\uparrow\uparrow\uparrow\updownarrow\updownarrow \updownarrow \updownarrow} \;,
\end{align} 
\ees
where the first (last) four spins are the core (ancillae) spins, and $\ket{\!\!\updownarrow}$ means that the spin can be either up or down. The all-spins down case \eqref{eq:alldown} results from the $ \ket{\!\!\downarrow\downarrow}$ configuration in Eq.~\eqref{2qspectrum}, while the $16$-fold degenerate case \eqref{eq:updown} results from the degeneracy of $\ket{\!\!\uparrow\uparrow}$ and$\ket{\!\!\downarrow\uparrow}$.

An important feature of the energy landscape of the $8$-qubit Hamiltonian is
that it does not have any local minima. This can be easily proved by
showing that a global minimum can always be reached from any state by
performing a sequence of single spin flips and never raising the
energy. To see this, consider an arbitrary state of the system. We can first flip
all the ancillae spins to $\ket{\!\!\downarrow}$ which, according to
Eq.~\eqref{2qspectrum}, can be done without raising the energy
(independently of the state of the corresponding core spin).  Then we
can flip the core spins in order to satisfy all the couplings between core spins, either
making them all $\ket{\!\!\downarrow}$ or all ${\!\!\ket \uparrow}$, whichever
requires the fewest spin flips.  Again, according to
Eq.~\eqref{2qspectrum}, this operation will not raise the energy of the
core-ancilla pair.
Hence, the final state is either the isolated ground state $\ket{\!\!\downarrow \downarrow \downarrow \downarrow \downarrow \downarrow
\downarrow \downarrow}$, or the state $\ket{\!\!\uparrow \uparrow
\uparrow \uparrow \downarrow \downarrow \downarrow \downarrow}$
that belongs to the degenerate cluster of ground state configurations.

\section{Simulated annealing and classical thermalization}

\begin{figure}[t!]
\begin{center}
\includegraphics [width= \columnwidth]{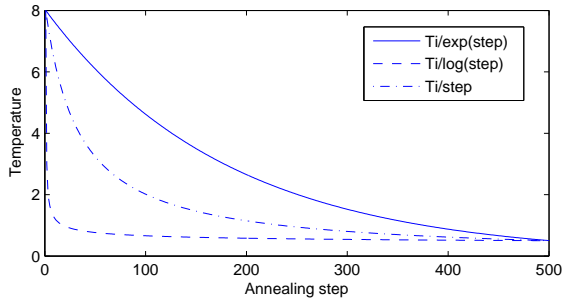}
\end{center}
\caption{Temperature as a function of annealing step $n$ for three different
schedules: exponential $T(n) = T_{i}r_{\exp}^n$, linear $T(n) =  T_{i}
/ (n r_{\rm lin}+1)$, and logarithmic $T(n) =  T_{i} /
(\log(n+1)r_{\log} +1)$, where $T_i$ is the initial temperature, $T_f$
is the final temperature, and $r_{\exp} = (T_{f}/T_{i})^{1/n_{\rm
tot}}$, $r_{\rm lin} = (T_{f}/T_{i}-1)/n_{\rm tot}$ and $r_{\log}
= (T_{f}/T_{i} -1) / \log(n_{\rm tot}+1)$, where $n_{\rm tot}$ is the total number of annealing steps.
}
\label{fig:SAschedules}
\end{figure}

\begin{figure}[t!]
\begin{center}
\includegraphics [width= \columnwidth]{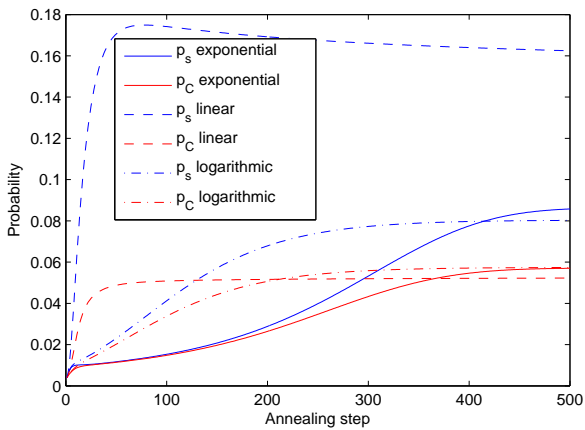}
\end{center}
\caption{Probabilities from SA for the three different
  schedules shown in Fig.~\ref{fig:SAschedules}. The probability of the isolated state $p_s$ is always
  higher than the average cluster state probability $p_C$.}
\label{fig:SAResultSchedules}
\end{figure}

Here we report the results of our numerical simulations of classical
thermalization (or SA), using the master
equation \eqref{eq:sa2}. In the plots below  we used
the Metropolis update rule for the transition probability $P(a \to a^{\pm}_{j})$. Explicitly, if $\Delta E = E_{a^{\pm}_{j}} - E_a$ is the energy
difference for the update, the transition probability is
\begin{align}
  \frac 1 {\textrm{ \# of spins}}  \min\left(1,\exp(-\beta \Delta E)\right)\;.
\end{align}
We have also tested other update rules, such as
Glauber's~\cite{Bertoin:2004uq}, and the results are essentially
unchanged. The result are also essentially unchanged when using a transition
probability $p \min\left(1,\exp(-\beta \Delta E)\right)$ for positive
$p$ (such as when simulating a continuous time master equation). We do
find a dependence on the choice of annealing schedule, i.e., the
functional dependence of the temperature on the number of steps. Three
different annealing schedules we used are shown in
Fig.~\ref{fig:SAschedules}, and the corresponding SA results are shown
in Fig.~\ref{fig:SAResultSchedules}. The
probability $p_s$ of the isolated state is always above the average
probability $p_C$ for a state in the cluster.

\begin{figure}[t!]
\begin{center}
\includegraphics [width= \columnwidth]{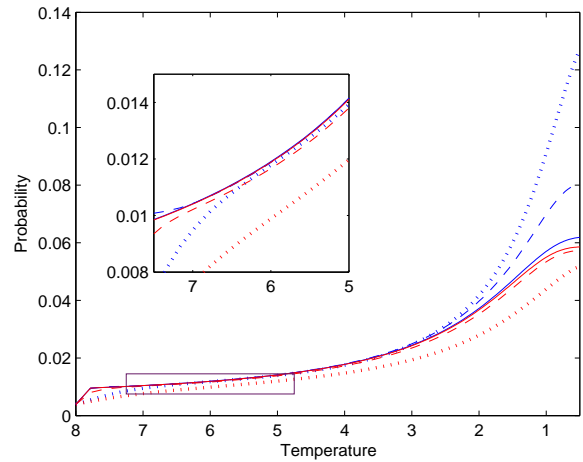}
\end{center}
\caption{Probabilities from SA for varying total numbers of
  steps. We used the Metropolis update rule with an exponential schedule. The
  lines correspond to 100 (dotted), 1000 (dashed) and 10000
  (solid) steps. The upper curves (blue) correspond to $p_s$, while the lower curves (red) correspond to $p_C$. The inset is a magnification of the boxed part. The separation between the probabilities of the isolated state and the cluster increases as the temperature decreases.}
\label{fig:SASeveralSteps}
\end{figure}

It might be argued that
thermalization at constant temperature corresponds most closely to the experimental situation, given that
the experimental system remains at an almost constant $17$mK.  This is modeled by the exponential annealing schedule, which rapidly converges to a nearly constant temperature, as can be seen in Fig.~\ref{fig:SAschedules}.  On the other hand, the energy scale of the Ising model changes during the QA evolution (see the
Fig.~2 insert in the main text), and the cooling schedule is determined not by the temperature alone but rather by the
ratio between the energy scale and the temperature.  We also show
$p_s$ and $P_C$ for an exponential schedule with Metropolis updates
and different numbers of steps in Fig.~\ref{fig:SASeveralSteps}.

\section{Classical master equation explanation for the enhancement of the isolated state}

We now explain why, as seen in the numerical simulations shown in Figs.~\ref{fig:SAResultSchedules} and \ref{fig:SASeveralSteps}, the probability of the isolated
state never exceeds that of the
average of the $16$ cluster ground states, i.e., why
\begin{align}
  p_s \ge \frac 1 {16} \sum_{i=1}^{16} p_i\;.
\end{align}
We are interested in sufficiently slow thermalization processes
(relative to spin flip rates), so that states connected by single spin-flips have similar populations.  The cluster of $16$ degenerate ground states and the isolated ground state are connected via a plateau of excited states
with energy $-4$. 

Let us first derive a rate equation for the isolated state. A single
spin-flip of a core spin in the isolated state raises its energy by
$4$, 
since it violates two couplings between the core spins and corresponds to a transition from $\ket{\!\!\downarrow\downarrow}$ to $\ket{\!\!\uparrow\downarrow}$ (where the second, ancilla, spin is unchanged), which doesn't change the energy according to Eq.~\eqref{2qspectrum}.
Likewise, a
single spin-flip of an ancilla spin in the isolated state violates no
couplings and corresponds to a transition from
$\ket{\!\!\downarrow\uparrow}$ to $\ket{\!\!\downarrow\uparrow}$ (with the core spin unchanged),
which raises the energy by $4$ according to
Eq.~\eqref{2qspectrum}. There are $8$ ways this can happen ($4$ core
and $4$ ancilla spins can be flipped). Since this accounts for all the single
spin transitions, Eq.~\eqref{eq:sa2} yields the rate equation
\begin{align}\label{eq:isoRate}
  \dot p_s &= 8 f(-4) p_e - 8 f(4) p_s \;,
\end{align}
where $p_e$ is the population of the excited states with energy
$-4$. Here we are assuming that the spin flip rate is the same for all
sites [corresponding to assuming $g^{(r)}_j = g^{(r)}$ in Eq.~\eqref{eq:SB}].

We next derive the rate equation for the cluster, once again accounting only for single spin flips. For states in the cluster the core spins are all up, and ancilla-spin flips are energy-preserving
transitions between states in the cluster. For core-spin flips we need to analyze two different situations. The first is a configuration in a ground state where the core-ancilla pair starts as $\ket{\!\!\uparrow\uparrow}$ and the core spin flips, so the state becomes $\ket{\!\!\downarrow\uparrow}$. This violates two couplings, with energy cost $4$, and according to Eq.~\eqref{2qspectrum} the energy difference between these two states is $4$, so the overall result is an excited state with energy $0$. The second is a configuration in a ground state where the core-ancilla pair starts as $\ket{\!\!\uparrow\downarrow}$ and again the core spin flips, so the state becomes $\ket{\!\!\downarrow\downarrow}$. This again violates two couplings, with energy cost $4$, but costs no energy according to Eq.~\eqref{2qspectrum}, so the overall result is an excited state with energy $-4$. 

Thus, a state with $l$ ancillae with spin
$\downarrow$ and $4-l$ ancillae with spin $\uparrow$ connects (via single spin-flips) to $l$
excited states with energy $-4$ and $4-l$ excited states with energy
$0$. To write a rate equation for ${p}_C = \sum_{i=1}^{16} p_i/16$ we shall
assume that all excited states with energy $0$ ($-4$) have probability $p(0)$ ($p_e$), and all
states in the ground state cluster have equal probability $p_C$. Summing
over the number $l$ of ancilla with spin $\downarrow$ for each cluster state,
the rate equation is
\bes
\begin{align}
  \sum_{i=1}^{16} \dot p_i &= \sum_{l = 0}^4 \binom 4 l \big(l f(-4)
  \,p_e - l f(4) \,p_C\nonumber \\
  &\qquad + (4-l) f(-8) \,p(0) - (4-l) f(8) \, p_C  \big)\,\\
  & = 32 \big( f(-4)\, p_e - f(4) \, p_C  \nonumber \\ & \qquad+
  f(-8) \, p(0) - f(8) \, p_C \big)\;,
\end{align}
\ees
so that
\beq
  \dot{p}_C  = 2 \big( f(-4)\, p_e - f(4) \, p_C + f(-8) \, p(0) - f(8) \, p_C \big) \;.
\eeq

For most temperatures of interest, relative to the energy scale of the
Ising Hamiltonian, the dominant transitions are those between the cluster and states with energy -4.  Transitions to energy 0 states are suppressed by the high energy cost, and transitions from energy 0 states to the cluster are suppressed by the low occupancy of the 0 energy states.
Then
\begin{align}\label{eq:coreRate}
      \dot{p}_C\approx 2 f(-4)\, p_e -2 f(4) \, p_C\;.
\end{align}

To show that $p_s \ge p_C$, assume that this is indeed the case initially. Then, in order for $p_C$ to become larger than $p_s$, they must first become equal at some inverse annealing temperature $\upbeta'$:  $p_s (\upbeta')= p_C (\upbeta') \equiv p_g$, and it suffices to check that this implies that $p_s$ grows faster than $p_C$. Subtracting Eq.~\eqref{eq:coreRate} from Eq.~\eqref{eq:isoRate} yields
\begin{align}
 \dot p_s- \dot p_C &= 6 \( f(-4)\, p_e - f(4) \, p_g\) \notag \\
 &= 6 f(-4) p_g \( \frac{p_e}{p_g} - \frac{P(g \to e)} {P(e \to g)}\)\;,
\end{align}
where in the second line we used Eq.~\eqref{eq:dbc}. Now, because the dynamical SA process we are considering proceeds via cooling, the ratio between the non-equilibrium excited state and
 the ground state probabilities will not be lower than the corresponding thermal equilibrium transition ratio, i.e., $\frac{p_e}{p_g} \geq  \frac{P(g \to e)} {P(e \to g)} = e^{-4\upbeta'}$. Therefore, as we set out to show,
 \begin{align}
     \dot p_s - \dot p_C \ge 0\;,
 \end{align}
 implying that at all times $p_s \ge p_C$.

\vspace{0.5cm}

\section{Degenerate perturbation theory explanation for quantum suppression of the isolated state}

We can understand the splitting of the degenerate ground subspace of the
Ising Hamiltonian $H_\is$ by treating the transverse field
$H_{\textrm{trans}} = -\sum_{j=1}^8 \sigma_j^x$ as a perturbation of the Ising
Hamiltonian $H_\is$ (thus treating the QA evolution as that of a closed system evolving backward in time). According to standard degenerate perturbation theory, the
perturbation $P_g$ of the ground subspace is given by the spectrum of the
projection of the perturbation on the ground subspace.  Denoting by
\beq
\Pi_0 = (\ket{\!\!\downarrow}\bra{\downarrow\!\!})^{\otimes 8} + \sum\ket{\!\!\uparrow\uparrow\uparrow\uparrow\updownarrow\updownarrow \updownarrow \updownarrow}\bra{\uparrow\uparrow\uparrow\uparrow\updownarrow\updownarrow \updownarrow \updownarrow\!\!}
\eeq 
the projector on the $17$-dimensional ground subspace, we therefore wish to understand the
spectrum of the operator
\begin{align}
P_g =  \Pi_0 \(-\sum_{j=1}^8 \sigma_j^x \)\Pi_0 \;.
\end{align}

The isolated state is  unconnected via single spin flips to any other state in the ground subspace, so we can write this operator as a direct sum of $0$ acting on the isolated state and the projection on the space $\Pi_0' = \Pi_0 - (\ket{\!\!\downarrow}\bra{\downarrow\!\!})^{\otimes 8} $ of the cluster
\begin{align}
P_g =  -0 \oplus \Pi_0' \(-\sum_{j=1}^8 \sigma_j^x \)\Pi_0' \;.
\end{align}
While $\sigma^x$ acting on any of the four ancillae
connects two cluster ground states, $\sigma^x$ acting on any core spin of a cluster state is projected away. Therefore the perturbation is given by the
operator
\begin{align}
P_g =  - 0 \oplus \(-\sum_{j=5}^8 \sigma_j^x\)\;,
\end{align}
where the sum is over the four ancillae spins.

Denoting the eigenbasis of $\sigma^x$ by $\ket{\pm} = (\ket{\!\!\uparrow}\pm\ket{\!\!\downarrow})/\sqrt{2}$, with respective eigenvalues $\pm 1$, the transverse field
splits the ground space of $H_{\is}$ lowering the energy of
$\ket{\!\!\uparrow\uparrow\uparrow\uparrow\!++++}$, and the four permutations of $\ket{-}$ in the ancillae spins of $\ket{\!\!\uparrow\uparrow\uparrow\uparrow\!+++-}$. None of these states
overlaps with the isolated ground state, which is therefore not a ground
state of the perturbed Hamiltonian. Furthermore, after the
perturbation, only the sixth excited state overlaps with the isolated state. The isolated state becomes a ground state only at the very end of the evolution (with time going forward), when the perturbation has vanished.

\section{The quantum Singular Coupling Limit does not agree with the experimental results}
Interestingly, an open system QA master equation in the singular coupling limit (SCL) yields results in qualitative agreement with classical thermalization, and opposite to our weak coupling limit (WCL) master equation \eqref{eq:H_S+H_LS}. Here, following Ref.~\onlinecite{Breuer:2002}, we present a derivation of the SCL master equation.

We consider a Hamiltonian of the form:
\begin{equation}
H(t) = H_S(t) + \epsilon^{-1} H_{I} + \epsilon^{-2} H_B \ ,
\end{equation}
where we take the interaction Hamiltonian $H_I$ to have the form  $A \otimes B$, where the system ($A$) and bath ($B$) operators are both Hermitian.  The formal solution in the interaction picture generated by $H_S$ and $H_B$ is given by:
\begin{equation}
\tilde{\rho}(t) = \tilde{\rho}(0) - i \epsilon^{-1} \int_0^t ds \left[ \tilde{H}_I(s), \tilde{\rho}(s) \right] \ .
\end{equation}
Plugging this solution back into the equation of motion and taking the partial trace over the bath, we obtain:
\begin{equation}
\frac{d}{dt} \tilde{\rho}_S(t) =- \epsilon^{-2} \int_0^t d s \mathrm{Tr}_B \left( \left[ \tilde{H}_I(t) , \left[ \tilde{H}_I(s), \tilde{\rho}(s) \right] \right] \right) \ ,
\end{equation}
where we have assumed that ${\textrm{Tr}}[\rho_B B] \equiv \langle B \rangle = 0$.  Under the standard Markovian assumption that $\rho(t) = \rho_S(t) \otimes \rho_B$ and under a change of coordinates $s = t - \tau$, we can write:
\begin{eqnarray}
\frac{d}{dt} \tilde{\rho}_S(t) &=&  \epsilon^{-2} \int_0^t d \tau  \left[\left(A(t) \tilde{\rho}(t-\tau) A(t-\tau) \right. \right. \\
&& \left. \left. - A(t) A(t- \tau) \tilde{\rho}(t- \tau) \right)\langle B(\tau) B(0) \rangle + \mathrm{h.c.} \right] \nonumber 
\end{eqnarray}
where $A(t) = U_S(t)AU_S^\dagger(t)$ and where we have used the homogeneity of the bath correlation function to shift its time-argument.
%
%
We change coordinates $\tau = \epsilon^2 \tau'$ and observe that under this coordinate change $\langle B(\tau) B(0) \rangle$ is independent of $\epsilon$.  We assume that this bath correlation function decays in a time $\tau_B$ that is sufficiently fast, such that $\tau_B \ll t / \epsilon^2$.  This allows us to approximate the integral by sending the upper limit to infinity.  We also assume that $\tau_B \ll \tau' \epsilon^2$, which forces the correlation time of the bath to zero, hence its spectral density to become flat, and hence---using the KMS condition---amounts to taking the infinite temperature limit.  Under these assumptions, we can now take the $\epsilon \to 0$ limit, yielding the SCL master equation
\begin{eqnarray}
\frac{d}{dt} {\rho}_S(t) &=&  - i \left[ H_S(t) + H_{\textrm{LS}}, \rho(t) \right] \nonumber   \\
&& + \gamma(0) \left( A  {\rho}(t)  A - \frac{1}{2} \left\{ A^2, \rho(t) \right\} \right) \ ,
\end{eqnarray}
where
\begin{eqnarray}
\gamma(\omega) &=& \int_{-\infty}^{\infty} d \tau' e^{- i \omega \tau'} \langle B(\tau') B(0) \rangle \ , \\
H_{\textrm{LS}} &=&  - A^2 \int_{-\infty}^{\infty} d \omega \gamma(\omega) \mathcal{P} \left( \frac{1}{\omega} \right) \ ,
\end{eqnarray}
where $H_{\textrm{LS}}$ is the Lamb shift (renormalization of the
system Hamiltonian) and where $\mathcal{P} $ denotes the Cauchy
principal value. Thus, even if $H_S$ is time-dependent, we recover the
same form for the SCL master equation as in the time-independent case~\cite{Breuer:2002}. This SCL master equation yields results in
qualitative agreement with classical thermalization, and opposite to
the WCL presented in Fig.~6 of the main text. Namely, the SCL master equation predicts that the isolated state is enhanced relative to the cluster of states. Since the bath and system-bath coupling dominate the system Hamiltonian in the SCL, it predicts that decoherence takes place not in the instantaneous energy eigenbasis of the system but in the computational basis.   Furthermore, the resulting Lamb shift in this limit, diagonal in the computational basis, preferentially lowers the energy of the all-up and all-down states relative to the other computational states.  Together, these two effects cause the isolated state to be more populated than the average population of the cluster at the end of the evolution, in contradiction to our experimental findings.  The agreement we find between our experimental results and the WCL master equation instead supports the idea that decoherence takes place in the instantaneous energy eigenbasis of the system and/or that the Lamb shift does not dominate the system Hamiltonian.

\section{Entanglement}
\begin{figure}[h]
\begin{center}
\includegraphics [width= \columnwidth]{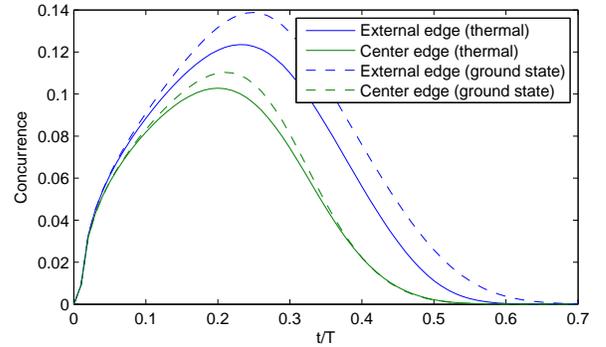}
\end{center}
\caption{Concurrence generated during our QA simulations, between a pair of ancillae qubits (``external edge") and a pair of core qubits (``center edge"). We show the concurrence for the
ground state and for the Gibbs state (``thermal"). Our master equation~\eqref{eq:H_S+H_LS} for the time-dependent density matrix gives concurrence values between these two extremes, that depend on the system-bath coupling strength used in the simulation.}
\label{fig:concurrences}
\end{figure}

Another interesting question is the possible presence of entanglement during the evolution. Answering this question experimentally requires measurements to be performed during the annealing, a capability that is absent from the device used in our experiments. However, when we compute the concurrence of the states obtained from our master equation~\eqref{eq:H_S+H_LS} during the QA evolution, which is consistent with the statistics of the measured output, we find it to be finite, as seen in Fig.~\ref{fig:concurrences}. This suggests that entanglement is being generated in our experiments.
  
\bibliography{experimentalQA}

\vspace{.5cm}

\end{document}